\documentclass{article}
\usepackage{spconf,amsmath,graphicx}
\usepackage{xcolor}
\usepackage{booktabs}
\usepackage{multirow}
\usepackage{url}

\title{Audio-Text Models Do Not Yet Leverage Natural Language}
\name{Ho-Hsiang Wu$^{1, \star}$, Oriol Nieto$^{2}$, Juan Pablo Bello$^{1, \dagger}$, Justin Salamon$^{2}$\thanks{$^{\star}$ Work done during an internship at Adobe Research.}
\thanks{$\dagger$ This work is partially supported by the National Science Foundation under award \#1955357}
}
\address{$^{1}$ Music and Audio Research Laboratory, New York University, New York, NY, USA \\
         $^{2}$ Adobe Research, San Francisco, CA, USA}

\begin{document}
\ninept
\maketitle
\begin{abstract}
Multi-modal contrastive learning techniques in the audio-text domain have quickly become a highly active area of research. Most works are evaluated with standard audio retrieval and classification benchmarks assuming that (i) these models are capable of leveraging the rich information contained in natural language, and (ii) current benchmarks are able to capture the nuances of such information. In this work, we show that state-of-the-art audio-text models do not yet really understand natural language, especially contextual concepts such as sequential or concurrent ordering of sound events. Our results suggest that existing benchmarks are not sufficient to assess these models' capabilities to match complex contexts from the audio and text modalities. We propose a Transformer-based architecture and show that, unlike prior work, it is capable of modeling the sequential relationship between sound events in the text and audio, given appropriate benchmark data.
We advocate for the collection or generation of additional, diverse, data to allow future research to fully leverage natural language for audio-text modeling.
\end{abstract}
\section{Introduction}
\label{sec:intro}

Multi-modal contrastive learning such as Contrastive Language-Image Pre-Training (CLIP) \cite{radford2021learning} has shown great success in various applications. In particular, it unlocks solutions to cross-modal tasks such as zero-shot image recognition \cite{zhai2022lit} or visual question answering \cite{lu2019vilbert, tan2019lxmert} in vision-language. 
Audio-\emph{visual} contrastive models have been used for localizing visual sound \cite{chen2021localizing, wu2022listen}, cross-modal retrieval \cite{suris2022s}, and zero-shot classification \cite{wu2022wav2clip, guzhov2022audioclip}. Recently, audio-\emph{text} models have received growing attention, as evidenced by the DCASE challenge on Language-Based Audio Retrieval \cite{xie2022language}, and have been applied to music audio for genre classification and tagging \cite{manco2022contrastive}, as well as environmental sounds for language-based audio retrieval \cite{zhao2021connecting, lou2022audio, mei2022metric, koepke2022audio}, and zero-shot classification tasks \cite{zhao2021connecting, elizalde2022clap}.

\begin{figure}[tb]
\centering
\includegraphics[width=\linewidth]{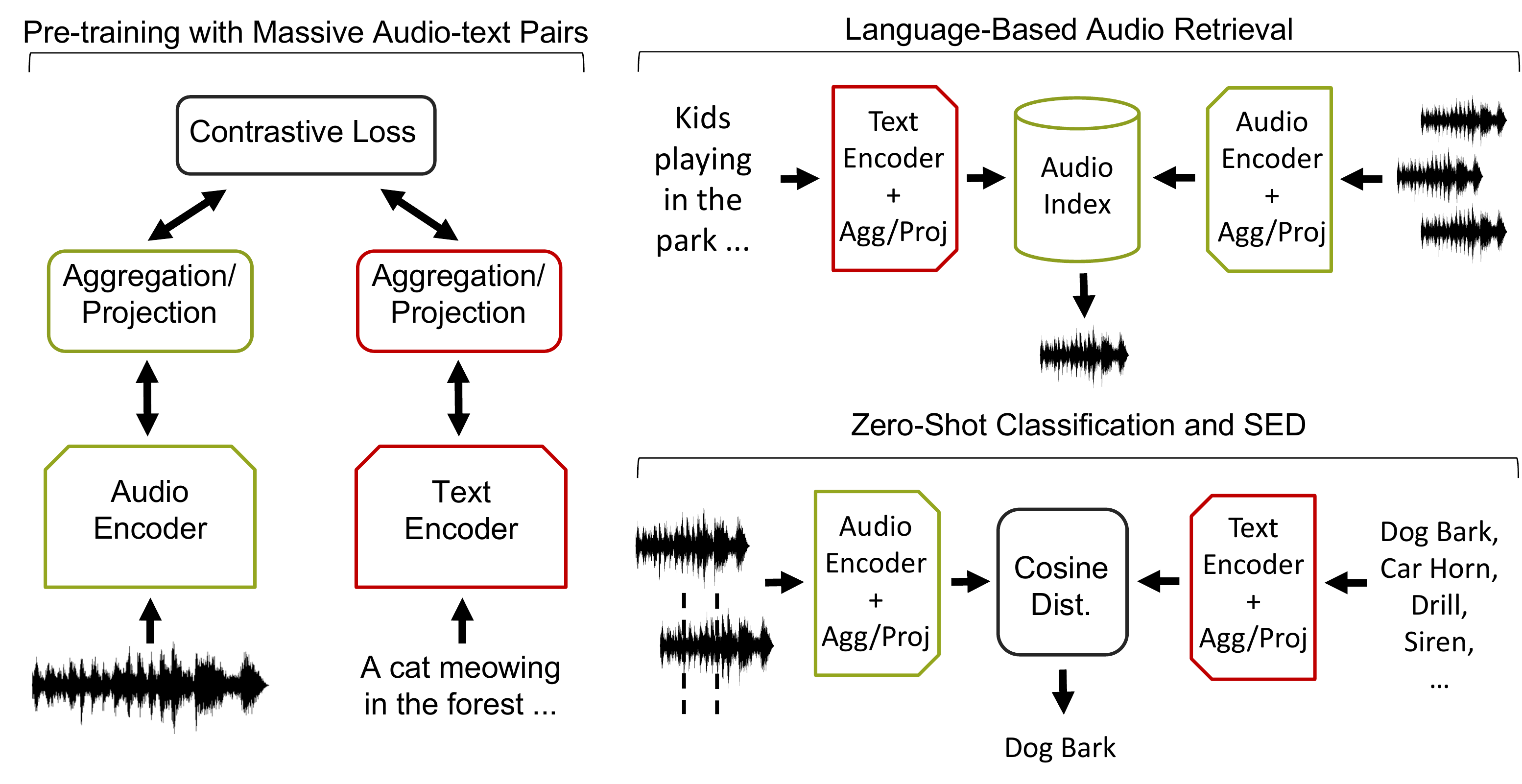}
\caption{Block diagrams for pre-training and downstream tasks.}
\label{fig:methods}
\end{figure}

Current state-of-the-art (SOTA) audio-text models utilize dual encoder architectures adopting popular pre-trained language (e.g., BERT \cite{devlin2018bert}) and audio (e.g., PANNs \cite{kong2020panns}) models to first encode each modality separately. 
Both encoders are concatenated with aggregation/projection (Agg/Proj) layers, and further trained with contrastive pretext tasks using audio-text pairs to learn to match between audio and natural language, as depicted in Figure \ref{fig:methods} on the left. 
Most prior works focus on exploring architectural choices such as encoders and Agg/Proj layers \cite{lou2022audio} or loss functions \cite{mei2022metric}, and are evaluated with common benchmarks for retrieval \cite{lou2022audio, mei2022metric, koepke2022audio} or classification \cite{zhao2021connecting, elizalde2022clap}. 
It is typically assumed that these approaches are capable of leveraging natural language to guide learning, since training data may include contextual descriptions of sound events using words such as ``then'', ``before'', or ``after'' for sequential relations and ``with'' or ``as'' for simultaneity.

However, there is a lack of exploration of what these audio-text models really learn. 
More specifically, are complex contextual ideas described in the text, such as sequential and concurrent ordering, successfully captured by existing systems?
In the 2022 DCASE Challenge on Language-Based Audio Retrieval (Task 6B) \cite{xie2022language}, most state-of-the-art (SOTA) systems employ pre-trained PANNs audio encoders with mean or max pooling over time, followed by multi-layer perceptron (MLP) projection layers \cite{xu2022_t6b, mei2022_t6b}. 
We hypothesize that these Agg/Proj mechanisms are not sufficient to capture the contextual complexities that often appear in natural language. 
This idea is related to studies that have shown that language models do not always leverage word ordering \cite{pham2020out}.

In this work, we present experiments designed to investigate current SOTA audio-text models' capabilities: What information are they leveraging to match audio to natural language queries? Can they actually model complex relationships such as sequential and concurrent ordering? How do model design choices such as the language model used impact the performance of the model?
Our main findings and contributions are as follows: (i) We show that current models focus on nouns and verbs for retrieval, and do not fully utilize the entire sentence; 
(ii) We show that SOTA systems that use temporal pooling and MLP as Agg/Proj mechanisms cannot capture concurrent or sequential sound event relations; (iii) We propose a Transformer-based architecture 
and show that, unlike prior work, it is capable of capturing sequential relationships between sound events in the text and audio, given sufficient training data and an appropriate benchmark; (iv) We find that existing benchmarks are insufficient for evaluating the sequential modeling abilities of audio-text models.

\section{Methodology}

\subsection{Baseline model and Transformer-based model}

We use a similar setup to the top performing systems in the 2022 DCASE Challenge on Language-Based Audio Retrieval \cite{xie2022language} as our baseline. 
For the audio encoder, we use a ResNet38 with pre-trained weights from PANNs \cite{kong2020panns} (2048 output dimensions), followed by an MLP projection layer to produce 1024 dimensions. For the text encoder we use RoBERTa-Large, as a more powerful replacement for the BERT model used in a top-performing system from the DCASE challenge \cite{mei2022_t6b}. We validate this choice in the preliminary experiment presented in Section~\ref{sec:lm}.
The model is pre-trained with InfoNCE loss \cite{oord2018representation} using a \(10^{-5}\) learning rate, batch size of \(64\), with standard StepLR scheduler (\(step\_size=20\), \(gamma=0.1\)) and early stopping criteria, on two V100 GPUs.

One of our hypotheses is that SOTA models represented by this baseline, with temporal pooling aggregation followed by MLP projection, will struggle to capture the sequencing of sound events over time. To investigate this, we propose a new architecture that replaces these Agg/Proj layers with a 2-layer 2-head Transformer with positional encoding, one for the text branch and one for the audio branch.
Each Transformer takes the encoder embeddings of all timestamps and outputs 1024 dimensional vectors. We use the output from the first position (i.e., the [CLS] token) as the final representation. On the text branch the Transformer takes the word (token) embeddings from the language encoder as input, and on the audio branch we apply the audio encoder to consecutive 1 s chunks to produce a sequence of audio embeddings, and pass these to the Transformer.

\subsection{Audio-text model pre-training}

The model is trained on a standard contrastive pretext task using pairs of audio and natural language text. 
Similar to CLAP \cite{elizalde2022clap}, we combine the training sets from multiple audio captioning datasets to create a single, larger, training set. These datasets include (i) AudioCaps \cite{kim2019audiocaps}: we were able to obtain 49k pairs (out of 50k); (ii) Clotho \cite{drossos2020clotho}: 3839 audio clips with 5 captions each from the training subset provided; (iii) MACS \cite{martin2021ground}: 3930 audio clips with 5 captions each; and (iv) FSD50K \cite{fonseca2021fsd50k}: 36k pairs resulting from concatenating labels and descriptions of corresponding audio metadata to form natural language sentences. In total, we obtain 123k audio-text pairs
which constitute our pre-training set.

\subsection{Language-based audio retrieval}

To probe the model, we consider the downstream task of language-based audio retrieval. This task models the application in which the user queries the system using natural language to
search for audio assets (clips) in a collection, 
and the audio assets are ranked by the system based on their similarity to the query text.

Given the query text, the model must retrieve its corresponding audio clip. The task is evaluated using a dataset of text-audio pairs. The pre-trained audio-text model is used to extract audio embeddings for the entire audio corpus, and text embeddings for all corresponding text descriptions.
Then, each text description is used as a query, and its embeddings are compared against all audio embeddings.
The top $k$ similar audio clips in the embedding space are retrieved, as shown on the top right of Figure \ref{fig:methods}. We report Recall at 10 (R@10) as the evaluation metric.
We use the following test sets,
which contain audio-text pairs where the text is a natural language description of the audio, for evaluation in our experiments:

\noindent \textbf{AudioCaps} \cite{kim2019audiocaps}: A subset of AudioSet with annotated natural language captions. The test set includes 814 audio clips, each 10 s long with 5 captions. For evaluation, we use each caption independently.

\noindent \textbf{Clotho} \cite{drossos2020clotho}: The test set includes 1045 audio-text pairs, where the audio is 15-30 s long. 
For the 2022 DCASE Challenge a new test set was released with another 1k pairs, we refer to it as \textbf{Clotho 2022}.

In Section \ref{sec:limitations} we will present a series of carefully designed experiments in which we manipulate the text and audio data either in the pre-training set used to train the audio-text model or in the aforementioned test sets, in ways that shed light on the model's capabilities and what it is leveraging to perform language-based audio retrieval.

\subsection{Choice of language model}
\label{sec:lm}

\begin{table}[t]
\centering
\begin{tabular}{@{\hskip 0.0in}c@{\hskip 0.02in}|@{\hskip 0.02in}c@{\hskip 0.03in}c@{\hskip 0.03in}c@{\hskip 0.02in}|@{\hskip 0.02in}c@{\hskip 0.03in}c@{\hskip 0.03in}c@{\hskip 0.00in}}
\toprule
Language & \footnotesize AudioCaps & \footnotesize Clotho & \footnotesize Clotho 2022 & \footnotesize ESC-50 & \footnotesize US8K & \footnotesize DESED \\
Model & \multicolumn{3}{c@{\hskip 0.02in}|@{\hskip 0.02in}}{R@10} & \multicolumn{3}{c}{F1} \\
\midrule
BERT-Base & 0.786 & 0.486 & 0.437 & \textbf{0.822} & \textbf{0.769} & \textbf{0.622} \\
BERT-Large & 0.794 & 0.498 & 0.474 & 0.794 & 0.746 & 0.612 \\
RoBERTa-Base & 0.779 & 0.479 & 0.444 & 0.757 & 0.742 & 0.596 \\
RoBERTa-Large & \textbf{0.798} & \textbf{0.505} & \textbf{0.477} & 0.78 & 0.753 & 0.611 \\
\bottomrule
\end{tabular}
\caption{Task performance versus LM text encoder for the baseline.}
\label{tab:results_retrieval}
\vspace{-1.em}
\end{table}

To guide our choice of pre-trained language model (LM) for the text encoder, we run preliminary experiments to evaluate its impact 
on the downstream performance of the baseline model. 
We consider two LM architectures that are commonly used by top audio-text models from the DCASE challenge: BERT~\cite{devlin2018bert} and RoBERTa \cite{liu2019roberta}. The latter is considered an improved version of the former as it uses dynamic masking, is trained with more data and larger batch size. We evaluate two variants for each model, Base and Large, with approximately 100M and 350M parameters respectively, and output embedding dimensions of 768 and 1024, respectively. 
We use pre-trained weights from huggingface\footnote{\url{https://huggingface.co/}}, take the embedding of the [CLS] token as the input to the MLP projection layer, and produce embeddings with 1024 dimensions. We do not freeze the LM weights, rather, we fine-tune both the audio and text encoders when pre-training the audio-text model.

In addition to our main downstream task, we include two additional tasks:  zero-shot classification and sound event detection (SED). For the former, we extract text embeddings from ground-truth labels and assign each audio test sample to its closest label in audio-text embedding space.
We evaluate the task on
ESC-50 \cite{piczak2015esc} and UrbanSound8K \cite{salamon2014dataset}
using the datasets' original splits, and report the F1 metric averaged across folds.
For SED, we slide the model's audio encoder over the signal to obtain embeddings every 50 ms and compare them to the text embedding of each class label. We apply a threshold to decide if the class is active in each frame. We use a random 20/80 split of the evaluation set to get validation and test sets, and use the former to find the optimal threshold value in [0, 1] and the latter to compute the evaluation metrics. We use
the public evaluation set of DESED \cite{Turpault2019_DCASE},
and report the standard segment-based SED metrics for 1 s segments. 

The results 
are shown in Table~\ref{tab:results_retrieval}. 
We see that the larger more powerful RoBERTa Large performs best across the three language-based audio retrieval benchmarks. 
Interestingly, BERT-Base works best for zero-shot classification and SED, suggesting the choice of LM should be informed by the target downstream task. Since our focus in this work is on language-based audio retrieval, 
we use RoBERTa-Large as the text encoder 
in all subsequent experiments.

\section{Limitations of current audio-text models}
\label{sec:limitations}

\subsection{Does the model leverage natural language or keywords?}

The goal of audio-text models is to leverage the full information in the natural language text query, i.e., to go beyond keywords (tags). Do current models achieve this? Or do they still mostly rely on keywords, which are typically nouns and verbs (e.g., ``dog'', ``barking'').
To answer this,
we pre-process the sentences in the pre-training data to strip the text of everything that is not a noun or verb. We use spaCy\footnote{\url{https://spacy.io/}} with its provided POS tagger for filtering. For example, the original sentence ``A vehicle is passing through a forest road as birds chirp in the background" becomes ``Vehicle passing forest road birds chirp background." 
Note that we only pre-process the pre-training data, while the downstream evaluation texts are not manipulated.

In Table \ref{tab:nv} we present the results for the baseline model trained either on the original data (Original) or the pre-processed sentences containing only noun and verb tokens (NV). 
For all three benchmark datasets,
the model trained on nouns and verbs performs similarly or better than the model trained with full sentences. This result is surprising, and suggests the model pre-trained on natural language may not be fully capturing the nuances and concepts in the sentences, and is mostly relying on nouns and verbs to match for retrieval.
Another possibility is that since the evaluation datasets are limited 
to mostly simple descriptions of sound events, they do not fully put the model's language understanding to the test,
yielding the illusion that current SOTA models that use temporal averaging for aggregation and MLP projection layers successfully leverage natural language.

\subsection{Does the model capture event ordering?}

Another potential benefit of audio-text models is that they can understand complex queries in which one sound happens \emph{after} another sound, or some sounds happen \emph{at the same time}. But do current models really capture these notions of event ordering and simultaneity? 
To investigate this, we design experiments with manipulated sentences \cite{ribeiro2020beyond} to see whether event ordering and simultaneity, expressed in the text using prepositions such as ``before'', ``after', ``then'', and ``as'', are captured by the audio-text model~\cite{hendricks2018localizing}. 

To see if the model can capture simultaneity, we take the 1640 sentences from our test sets that contain ``as'' and replace the preposition with ``then'', in this way changing their meaning from describing concurrent sounds to describing consecutive sounds. We do the opposite for the 1205 test sentences that contain ``then'', replacing it with ``as''. We then use the manipulated text to query for the original audio, and report R@10 where the retrieval set is all other sentences with the same preposition. If the model can differentiate between concurrent and consecutive events, we should see a notable drop in performance for the manipulated sentences compared to using the original ones. The results are reported in Table~\ref{tab:results_then_as}, where were observe only a marginal performance change. This suggests the model cannot capture the concept of simultaneous sounds.

\begin{table}[t]
\centering
\begin{tabular}{c@{\hskip 0.13in}c@{\hskip 0.13in}c@{\hskip 0.13in}c}\toprule
& AudioCaps & Clotho & Clotho 2022 \\
\midrule
Original & \textbf{0.798} & 0.505 & 0.477  \\
NV & 0.796 & \textbf{0.524} & \textbf{0.506}  \\
\bottomrule
\end{tabular}
\caption{R@10 for the baseline audio-text model pre-trained using the original text data (Original) and the filtered text data that only contains nouns and verbs (NV).}
\label{tab:nv}
\end{table}

\begin{table}[t]
\centering
\begin{tabular}{cc|cc}\toprule
Then & r/Then/As & As & r/As/Then \\
\midrule
0.863 & 0.844 & 0.822 & 0.819 \\
\bottomrule
\end{tabular}
\caption{R@10 for the baseline model on sentences containing ``then'' or ``as'', evaluated against the original sentences and on the same sentence after swapping the two prepositions.}
\label{tab:results_then_as}
\end{table}

\begin{table}[t]
\centering
\begin{tabular}{
@{\hskip 0.01in}c|@{\hskip 0.03in}c@{\hskip 0.03in}c@{\hskip 0.03in}c@{\hskip 0.03in}c@{\hskip 0.03in}c@{\hskip 0.01in}}
\toprule
  & ``Before" & ``After" & ``Then" & ``Followed by" \\
\midrule
PTe & 185 (1.8\%) & 105 (1.0\%) & 1205 (11.7\%) & 1279 (12.4\%) \\
Tr & 1036 (0.8\%) & 1071 (0.9\%) & 8177 (6.6\%) & 8321 (6.8\%) \\
\bottomrule
\end{tabular}
\caption{
Top: Number of sentences in the preposition test set (PTe) containing each preposition, with the \% it represents of the full test-set in parenthesis. Bottom: the same distribution for the pre-training set (Tr), with the \% it represents of the pre-training set in parenthesis.}
\label{tab:preposition}
\end{table}

\begin{figure}[t]
  \centering
  \includegraphics[width=0.8\linewidth]{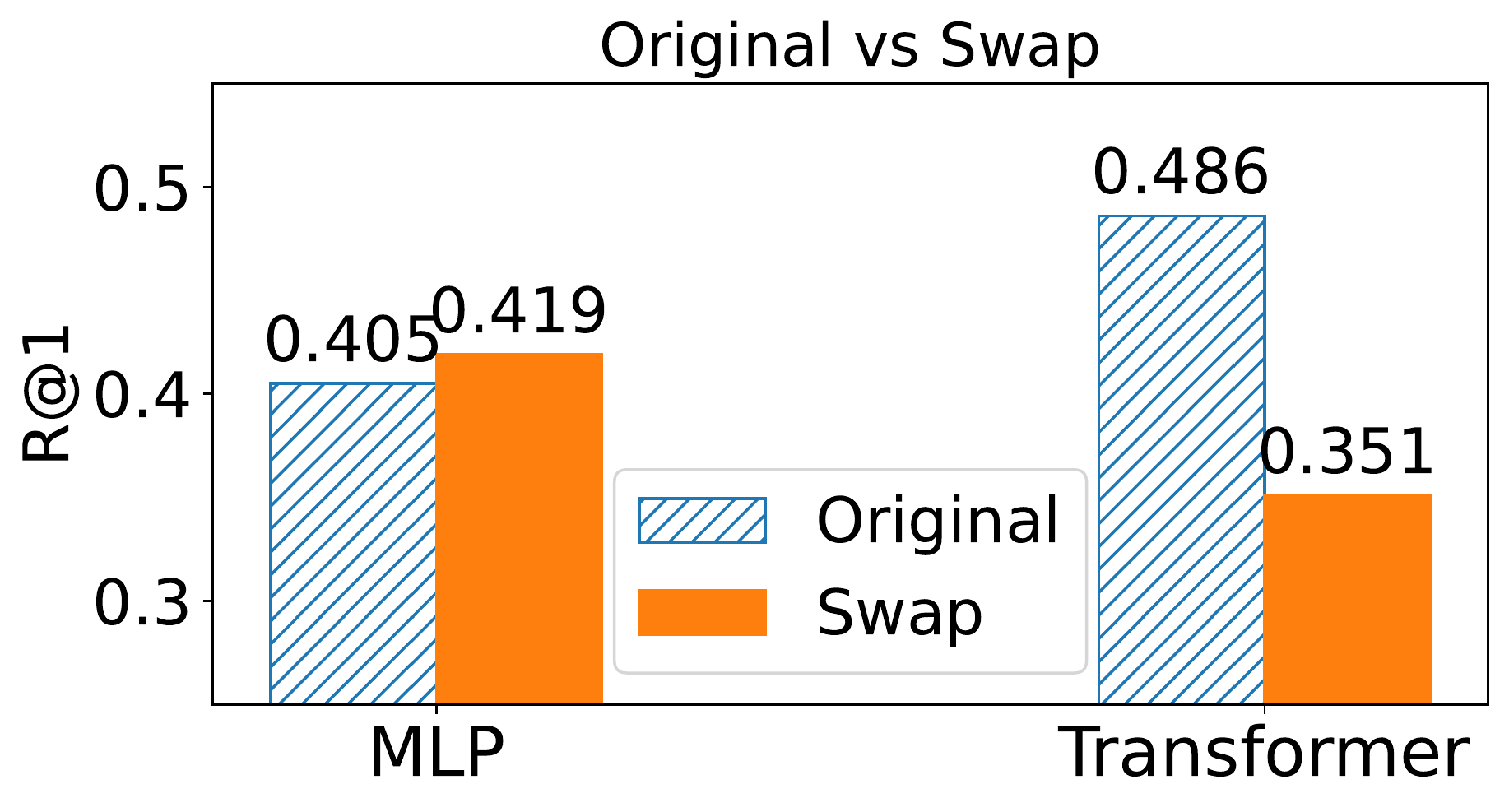}
    \caption{R@1 on the preposition test set (PTe) for the baseline (MLP) and our proposed model (Transformer). Lined bars show results using the original sentences in PTe as queries, and solid orange bars show results after swapping the clauses before/after the preposition.}
  \label{fig:swap}
\end{figure}

Next, we dig deeper into whether the model can capture event ordering.
We create a ``preposition test set'', PTe, by taking the sentences in the test subsets of AudioCaps, Clotho, and Clotho 2022, and keeping only sentences that contain one of the following prepositions: ``before'', ``after'', ``then'', and ``followed by''. The distribution of sentences with prepositions in PTe is provided in the top row of Table~\ref{tab:preposition}. In the bottom row we provide the distribution of sentences with prepositions in the pre-training set.
We then run the following experiment: we take PTe and swap the order of the clauses before and after the preposition to form new sentences. 
For example, ``Two cars drive past \emph{before} a distant semi truck honks'' becomes ``A distant semi truck honks \emph{before} two cars drive past''. 
We use either the original sentences or the swapped sentences as queries to retrieve the original audio, and report Recall at 1 (R@1) to examine the model under stringent conditions.
As with the previous experiment, if the model can capture sequential ordering, then its performance should drop notably for the swapped sentences compared to the original ones, since the manipulated texts list the sound events in reverse order to their true sequencing in the audio.
We compare the performance of the baseline model on the two query sets against our proposed Transformer-based model which, we hypothesize, should benefit from its ability to model temporal sequencing.
The results are shown in Figure \ref{fig:swap}, with the baseline and our proposed architecture indicated by their projection layer: MLP and Transformer, respectively.
The baseline performs equally (even marignally better) when we swap the order of events in the query text, indicating it is not capturing the ordering indicated by the prepositions.
Conversely, the Transformer, which performs better on the original sentences, drops 27 points for the swapped sentences, indicating it is capable of modeling the sequencing of sound events.

\begin{figure}[t]
  \centering
  \includegraphics[width=0.8\linewidth]{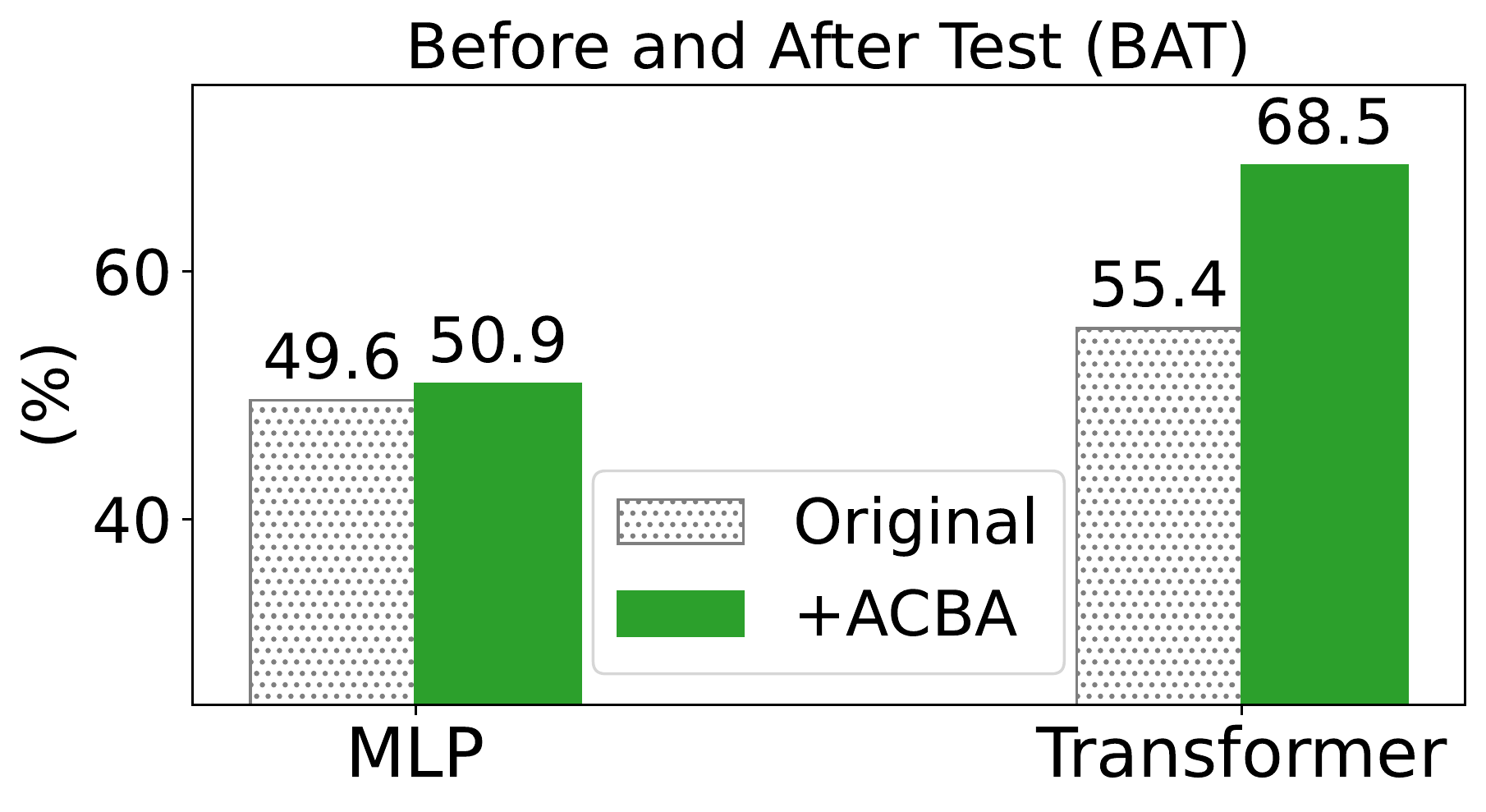}
  \caption{Performance on the Before and After Test (BAT) set for the baseline (MLP) and our proposed model (Transformer). Models trained either on the original pre-training set (dotted bars) or the augmented +ACBA set (solid green bars). Performance measured as \% of test samples for which the original text query is closer to the corresponding audio than the manipulated text (50\% = random guess).}
  \label{fig:bat}
\end{figure}

To further study the effect of prepositions, we run an additional experiment
where we take a balanced set of 176 sentences from PTe, half containing ``before" and half ``after'', 
referred to as Before After Test (BAT). 
We compute the embedding distance between the audio and these test sentences, with and without swapping the event order by replacing ``before'' with ``after'' and vice versa, and count the \% of times the original sentence is closer to the audio than the swapped one. 
For example, the sentence from the earlier example becomes ``Two cars drive past \emph{after} a distant semi truck honks." 
If the models fail to capture event ordering, the result should be roughly 50\% (i.e., random guessing).
The results are reported in Figure \ref{fig:bat} by the bars marked ``Original''.
The baseline (MLP) yields roughly 50\%, indicating it cannot capture the difference between ``before'' and ``after''. 
The transformer does better than random (55\%), albeit moderately.

\begin{table}[t]
\centering
\begin{tabular}{cc|ccc}
\toprule
Agg/Proj & Data & AudioCaps & Clotho & Clotho 2022 \\ %
\midrule
MLP & Original & \textbf{0.798} & \textbf{0.505} & \textbf{0.477} \\ %
Transformer & Original & 0.685 & 0.414 & 0.393 \\ %
\midrule
MLP & +ACBA & 0.790 & 0.500 & 0.473 \\ %
Transformer & +ACBA & 0.730 & 0.425 & 0.405 \\ %
\bottomrule
\end{tabular}
\caption{R@10 for the baseline model (MLP) and the proposed Transformer-based model (Transformer). Top: results when training with the original pre-training set (Original). Bottom: results when training on the augmented pre-training set which includes the additional 50k ACBA sentences.}
\label{tab:before_after}
\vspace{-1.0em}
\end{table}

\subsection{Are the training data sufficient to learn about ordering?}

Could the limited gains on BAT when using a Transformer be due to a lack of data?
Sentences with prepositions (PTe) represent less than 10\% of the pre-training data (Tr), and less than 1\% 
of the sentences in Tr
contain ``before'' or ``after''. To answer this, we create synthetic audio-text pairs to augment the data for these prepositions. 
We take the 19k sentences in the AudioCaps training set that do not already contain the prepositions in PTe and use them to create new sentences:
we randomly select two sentences, treat them as sub-clauses, and concatenate them with a preposition 
in between 
(e.g., ``\mbox{\textlangle{sentence} A}\textrangle~before \textlangle{sentence} B\textrangle''). 
We concatenate the audio clips associated with each sentence in the order indicated by the chosen preposition with a 1 s cross-fade.
We generate 50k such pairs\footnote{\url{github.com/hohsiangwu/preposition-synthesis}}, referred to as AudioCaps-BeforeAfter (ACBA), and create a new pre-training set (+ACBA) by adding them to the 123k existing pre-training pairs.

The results of pre-training the baseline (MLP) and our proposed model (Transformer) on +ACBA and evaluating them on BAT are plotted as solid green bars in Figure \ref{fig:bat}.
The performance of our Transformer model is notably boosted by 13 points, strengthening the evidence that it can model temporal ordering in the text and audio. It is also a clear indication that the existing pre-training data are insufficient for teaching audio-text models about sound event ordering.
Even with the additional ACBA sentences, the baseline (MLP) is still equivalent to a random guess (51\%), supporting our conjecture of an inherent limitation in the architecture.

Finally, we compare our Transformer-based architecture to the baseline (MLP) on the original language-based audio retrieval benchmarks, trained on either our original pre-training set or the augmented set +ACBA, presented in Table \ref{tab:before_after}. Surprisingly, in contrast to our prior experiments, the baseline outperforms the Transformer, even when we add the ACBA training data. One hypothesis is that the Transformer is still under trained: its performance improves with the addition of ACBA, so it is plausible it will improve further with more data. A second is that since sentences with prepositions represent a only small percentage of the test data in existing benchmarks, the benefits of the Transformer for sequence modeling do not come to light here.
This highlights the limitations of existing open datasets for training and evaluating audio-text models. With larger and more complex training sets, and test sets that require the model to leverage natural language to a greater extent, we think it is likely the Transformer will surpass the baseline, in accordance with our other experiments.

\section{Conclusion}
\label{sec:conclusion}

Contrastive learning for the audio and text modalities holds the promise of unlocking powerful applications such as language-based audio retrieval, zero-shot audio classification and open-vocabulary sound event detection.
In this work we presented several experiments aimed at understanding the capabilities of current top-performing audio-text models.
We showed that despite being trained on audio-text pairs where the text is natural language sentences, these architectures fail to fully leverage the information in the natural language signal: they cannot capture concepts such as simultaneity and event ordering, and models trained on text limited to nouns and verbs perform equally well as those trained on full sentences. To alleviate these limitations we proposed a Transformer-based architecture and showed that, given sufficient training data and a benchmark designed to probe the model's ability to capture event ordering, it outperforms the baseline considerably. Conversely, it does not outperform the baseline on the existing benchmarks, suggesting they are not well suited for understanding the audio-text model's abilities to capture more complex relationships between natural language and the audio signal. To reach the multi-modal understanding capabilities shown for other modalities 
\cite{radford2021learning}, we advocate for the creation of significantly larger datasets with ample representation  of audio-text pairs that describe complex relationships such as sound event simultaneity and 
sound event 
ordering.

\bibliographystyle{IEEEbib}
\bibliography{refs}

\begin{thebibliography}{10}

\bibitem{radford2021learning}
Alec Radford, Jong~Wook Kim, Chris Hallacy, Aditya Ramesh, Gabriel Goh,
  Sandhini Agarwal, Girish Sastry, Amanda Askell, Pamela Mishkin, et~al.,
\newblock ``Learning transferable visual models from natural language
  supervision,''
\newblock {\em ICML}, 2021.

\bibitem{zhai2022lit}
Xiaohua Zhai, Xiao Wang, Basil Mustafa, Andreas Steiner, Daniel Keysers,
  Alexander Kolesnikov, and Lucas Beyer,
\newblock ``Lit: Zero-shot transfer with locked-image text tuning,''
\newblock in {\em CVPR}, 2022, pp. 18123--18133.

\bibitem{lu2019vilbert}
Jiasen Lu, Dhruv Batra, Devi Parikh, and Stefan Lee,
\newblock ``Vilbert: Pretraining task-agnostic visiolinguistic representations
  for vision-and-language tasks,''
\newblock {\em NeurIPS}, vol. 32, 2019.

\bibitem{tan2019lxmert}
Hao Tan and Mohit Bansal,
\newblock ``Lxmert: Learning cross-modality encoder representations from
  transformers,''
\newblock {\em EMNLP}, 2019.

\bibitem{chen2021localizing}
Honglie Chen, Weidi Xie, Triantafyllos Afouras, Arsha Nagrani, Andrea Vedaldi,
  and Andrew Zisserman,
\newblock ``Localizing visual sounds the hard way,''
\newblock in {\em CVPR}, 2021, pp. 16867--16876.

\bibitem{wu2022listen}
Ho-Hsiang Wu, Magdalena Fuentes, Prem Seetharaman, and Juan~Pablo Bello,
\newblock ``How to listen? rethinking visual sound localization,''
\newblock {\em INTERSPEECH}, 2022.

\bibitem{suris2022s}
D{\'\i}dac Sur{\'\i}s, Carl Vondrick, Bryan Russell, and Justin Salamon,
\newblock ``It's time for artistic correspondence in music and video,''
\newblock in {\em CVPR}, 2022, pp. 10564--10574.

\bibitem{wu2022wav2clip}
Ho-Hsiang Wu, Prem Seetharaman, Kundan Kumar, and Juan~Pablo Bello,
\newblock ``Wav2clip: Learning robust audio representations from clip,''
\newblock in {\em ICASSP}, 2022.

\bibitem{guzhov2022audioclip}
Andrey Guzhov, Federico Raue, J{\"o}rn Hees, and Andreas Dengel,
\newblock ``Audioclip: Extending clip to image, text and audio,''
\newblock in {\em ICASSP}, 2022, pp. 976--980.

\bibitem{xie2022language}
Huang Xie, Samuel Lipping, and Tuomas Virtanen,
\newblock ``Language-based audio retrieval task in dcase 2022 challenge,''
\newblock {\em DCASE Workshop}, 2022.

\bibitem{manco2022contrastive}
Ilaria Manco, Emmanouil Benetos, Elio Quinton, and Gy{\"o}rgy Fazekas,
\newblock ``Contrastive audio-language learning for music,''
\newblock {\em ISMIR}, 2022.

\bibitem{zhao2021connecting}
Yanpeng Zhao, Jack Hessel, Youngjae Yu, Ximing Lu, Rowan Zellers, and Yejin
  Choi,
\newblock ``Connecting the dots between audio and text without parallel data
  through visual knowledge transfer,''
\newblock {\em NAACL}, 2022.

\bibitem{lou2022audio}
Siyu Lou, Xuenan Xu, Mengyue Wu, and Kai Yu,
\newblock ``Audio-text retrieval in context,''
\newblock in {\em ICASSP}. IEEE, 2022, pp. 4793--4797.

\bibitem{mei2022metric}
Xinhao Mei, Xubo Liu, Jianyuan Sun, Mark~D Plumbley, and Wenwu Wang,
\newblock ``On metric learning for audio-text cross-modal retrieval,''
\newblock {\em INTERSPEECH}, 2022.

\bibitem{koepke2022audio}
A~Sophia Koepke, Andreea-Maria Oncescu, Joao Henriques, Zeynep Akata, and
  Samuel Albanie,
\newblock ``Audio retrieval with natural language queries: A benchmark study,''
\newblock {\em IEEE Transactions on Multimedia}, 2022.

\bibitem{elizalde2022clap}
Benjamin Elizalde, Soham Deshmukh, Mahmoud~Al Ismail, and Huaming Wang,
\newblock ``Clap: Learning audio concepts from natural language supervision,''
\newblock {\em arXiv preprint arXiv:2206.04769}, 2022.

\bibitem{devlin2018bert}
Jacob Devlin, Ming-Wei Chang, Kenton Lee, and Kristina Toutanova,
\newblock ``Bert: Pre-training of deep bidirectional transformers for language
  understanding,''
\newblock {\em NAACL}, 2019.

\bibitem{kong2020panns}
Qiuqiang Kong, Yin Cao, Turab Iqbal, Yuxuan Wang, Wenwu Wang, and Mark~D
  Plumbley,
\newblock ``Panns: Large-scale pretrained audio neural networks for audio
  pattern recognition,''
\newblock {\em IEEE/ACM Transactions on Audio, Speech, and Language
  Processing}, vol. 28, pp. 2880--2894, 2020.

\bibitem{xu2022_t6b}
Xuenan Xu, Zeyu Xie, Mengyue Wu, and Kai Yu,
\newblock ``The {SJTU} system for {DCASE2022} challenge task 6: Audio
  captioning with audio-text retrieval pre-training,''
\newblock Tech. {R}ep., DCASE2022 Challenge, July 2022.

\bibitem{mei2022_t6b}
Xinhao Mei, Xubo Liu, Haohe Liu, Jianyuan Sun, Mark~D. Plumbley, and Wenwu
  Wang,
\newblock ``Language-based audio retrieval with pre-trained models,''
\newblock Tech. {R}ep., DCASE2022 Challenge, 2022.

\bibitem{pham2020out}
Thang~M Pham, Trung Bui, Long Mai, and Anh Nguyen,
\newblock ``Out of order: How important is the sequential order of words in a
  sentence in natural language understanding tasks?,''
\newblock {\em Findings of the ACL: ACL-IJCNLP}, 2021.

\bibitem{oord2018representation}
Aaron van~den Oord, Yazhe Li, and Oriol Vinyals,
\newblock ``Representation learning with contrastive predictive coding,''
\newblock {\em arXiv preprint arXiv:1807.03748}, 2018.

\bibitem{kim2019audiocaps}
Chris~Dongjoo Kim, Byeongchang Kim, Hyunmin Lee, and Gunhee Kim,
\newblock ``Audiocaps: Generating captions for audios in the wild,''
\newblock in {\em NAACL}, 2019, pp. 119--132.

\bibitem{drossos2020clotho}
Konstantinos Drossos, Samuel Lipping, and Tuomas Virtanen,
\newblock ``Clotho: An audio captioning dataset,''
\newblock {\em ICASSP}, 2020.

\bibitem{martin2021ground}
Irene Mart{\'\i}n-Morat{\'o} and Annamaria Mesaros,
\newblock ``What is the ground truth? reliability of multi-annotator data for
  audio tagging,''
\newblock in {\em EUSIPCO}. IEEE, 2021, pp. 76--80.

\bibitem{fonseca2021fsd50k}
Eduardo Fonseca, Xavier Favory, Jordi Pons, Frederic Font, and Xavier Serra,
\newblock ``Fsd50k: an open dataset of human-labeled sound events,''
\newblock {\em IEEE/ACM Transactions on Audio, Speech, and Language
  Processing}, vol. 30, pp. 829--852, 2021.

\bibitem{liu2019roberta}
Yinhan Liu, Myle Ott, Naman Goyal, Jingfei Du, Mandar Joshi, Danqi Chen, Omer
  Levy, Mike Lewis, Luke Zettlemoyer, and Veselin Stoyanov,
\newblock ``Roberta: A robustly optimized bert pretraining approach,''
\newblock {\em arXiv preprint arXiv:1907.11692}, 2019.

\bibitem{piczak2015esc}
Karol~J Piczak,
\newblock ``Esc: Dataset for environmental sound classification,''
\newblock in {\em ACM Multimedia}, 2015, pp. 1015--1018.

\bibitem{salamon2014dataset}
Justin Salamon, Christopher Jacoby, and Juan~Pablo Bello,
\newblock ``A dataset and taxonomy for urban sound research,''
\newblock in {\em ACM Multimedia}, 2014, pp. 1041--1044.

\bibitem{Turpault2019_DCASE}
Nicolas Turpault, Romain Serizel, Ankit Parag~Shah, and Justin Salamon,
\newblock ``{Sound event detection in domestic environments with weakly labeled
  data and soundscape synthesis},''
\newblock in {\em {DCASE}}, New York City, United States, October 2019.

\bibitem{ribeiro2020beyond}
Marco~Tulio Ribeiro, Tongshuang Wu, Carlos Guestrin, and Sameer Singh,
\newblock ``Beyond accuracy: Behavioral testing of nlp models with checklist,''
\newblock {\em ACL}, 2020.

\bibitem{hendricks2018localizing}
Lisa~Anne Hendricks, Oliver Wang, Eli Shechtman, Josef Sivic, Trevor Darrell,
  and Bryan Russell,
\newblock ``Localizing moments in video with temporal language,''
\newblock {\em EMNLP}, 2018.

\end{thebibliography}

\end{document}